\documentclass[hidelinks]{interact}
\usepackage{tikz,textpos}
\usepackage{epstopdf}
\usepackage{subfigure}

\usepackage{natbib}
\bibpunct[, ]{(}{)}{;}{a}{}{,}

\theoremstyle{plain}

\theoremstyle{definition}

\theoremstyle{remark}

\begin{document}


\title{A State-Space Model for Assimilating Passenger and Vehicle Flow Data with User Feedback in a Transit Network}

\author{
\name{S. Arabas and A.E. Papacharalampous}
\thanks{Contact: A.E. Papacharalampous. Email: a.papacharalampous@aethon.gr}
\thanks{Paper presented at the {\em Mathematics Applied in Transport and Traffic Systems}, TUDelft, October 2018}
\thanks{~~~~~~~~~~~This project has received funding from the European Union's Horizon 2020}
\thanks{~~~~~~~~~~~research and innovation programme under grant agreement No 739607.}
\affil{AETHON Engineering Consultants, Athens, Greece}
}

\maketitle 
\begin{abstract}
This note explores the idea of utilising a state-space model, congruent with the underlying equations 
  of the Kalman filter with control input, for reconstructing the state of crowdedness in a transit network.
The envisaged role of the proposed scheme is twofold:
  first, to provide an estimate of the state of crowdedness given input data on vehicle movement, 
  on passenger inflow/outflow at stations and on measured crowdedness; 
  second, to trigger localised requests for feedback based on the estimated system state as well as on the data assimilation performance indices. 
The latter is applicable to a scenario where the crowdedness is measured through passenger feedback. 
The feedback loop is conceptualised to be realised with a participatory crowd-sensing smartphone-based system in which  
  reported perceived levels of crowdedness are assimilated in near-real-time with the aim of improving the estimation of the model state. 
Presented model is also applicable for assimilating other relevant measurements, for instance, vehicle weighing,
  automatic passenger counting, aggregated smartcard data or passive wireless device monitoring data.
\end{abstract}


\begin{textblock*}{20pt}(-3pt,226pt)
\begin{tikzpicture}[scale=.018]
\fill[fill={rgb,255:red,0;green,51;blue,153}] (-27,-18) rectangle (27,18);  
\pgfmathsetmacro\inr{tan(36)/cos(18)}
\foreach \i in {0,1,...,11} {
  \begin{scope}[shift={(30*\i:12)}]
    \fill[fill={rgb,255:red,255;green,204;blue,0}] (90:2)
        \foreach \x in {0,1,...,4} { -- (90+72*\x:2) -- (126+72*\x:\inr) };
  \end{scope}
}
\end{tikzpicture}
\end{textblock*}
\vspace{-25pt}
\section{Introduction}
\vspace{-5pt}

Despite ubiquitous and ever-growing availability of 
  datasets on the operation and schedules of public transportation,
  crowdedness levels -- to the authors' knowledge -- are currently
  not among the variables disseminated through public channels.
Crowdedness is intuitively one of the key factors affecting 
  perceived level of service what has been corroborated in 
  numerous research reports \citep[see e.g.,][and references therein]{Haywood_et_al_2017}.
The level of crowding determines the ability for a passenger to seat,
  the ability to spend time productively on board, and -- in the extreme case -- the ability to board.
Moreover, high passenger density causes stress due to safety, security,
  hygiene or thermal comfort concerns, and all of these have the potential 
  to influence route and mode choices of passengers \citep[see e.g.,][]{Kim_et_al_2015}.
From the operators' perspective, overcrowding translates to
  increased accident risk, decreased customer satisfaction, and suboptimal infrastructure utilisation 
  due to lengthening of dwell times and unbalanced passenger loading potentially leading to unused seating capacity
  \citep[see][for examples of modelling and data-analysis studies, and a review, respectively]{Regirer_and_Shapovalov_2003,Ball_2016,Haywood_et_al_2017}.

The stimuli for the presented work have been the realisations that: 
  (i) given even a simple macroscopic model of passenger loading dynamics,
  the levels of crowdedness are retrievable from existing datasets 
  \citep[as demonstrated e.g., in][]{Reyes_and_Cipriano_2014} and 
  (ii) crowdedness is one of the variables that is ``measurable''
  through passenger feedback \citep[e.g.,][]{Farkas_et_al_2015} 
  realised with a participatory crowd-sensing smartphone-based system 
  \citep{Guo_et_al_2016}.
Achieving both of the above would create the potential for enrichment of existing public 
  transportation datasets with assimilated levels of crowdedness.
This, in turn, creates the potential for incorporation of information on crowdedness 
  in route planning \citep[e.g.,][]{Handte_et_al_2016} 
  as well as in real-time provision of data to passengers \citep[e.g.,][]{Nuzzolo_et_al_2016, Drabicki_et_al_2017},
  all with the aim of optimising network usage and passenger comfort.

The focus of this work is placed on formulation and demonstration of a simple 
  data fusion methodology capable of assimilation of the following data into a macroscopic simulation of 
  passenger distribution in a transit network:
  (i)~data on passenger inflows/outflows to/from stations (differential macroscopic data), 
  (ii)~measurements of passenger density (integral macroscopic data),
  (iii)~vehicle location information (microscopic data). 

The concept of a feedback loop between the system and its users is conceptualised
  similarly as in \citet{Farkas_et_al_2015} with a simple yet real-time-processed
  smartphone interface for reporting perceived crowdedness around a user.
Noteworthy, interpreting user perceptions of crowdedness is a complex problem
  that has been researched over the years from psychological/behavioural angles 
  \citep{Stokols_1972,Kalb_and_Keating_1981,Mohd_Mahudin_et_al_2012,Li_and_Hensher_2013}.
In a wider context, human perception of numerosity is a factor here with
  suggested geometric or logarithmic (Weber-Fechner-like) principles
  \citep{Toyosawa_and_Kawai_2005,Cicchini_et_al_2016}.
In the present study, from the point of view of the data assimilation
  approach, such ``measurements'' are treated as any other noisy data
  on crowdedness and the methodology is, in principle,
  applicable for assimilation of data from vehicle weighing
  \citep{Frumin_2010,Nielsen_et_al_2014,Ball_2016},
  data obtained by counting distinct personal electronic devices
  \citep[e.g.,][]{Schauer_et_al_2014}, data available in smartcard-based
  ticketing systems \citep[e.g.,][]{Zhang_et_al_2017} or data from
  automatic passenger counters \citep[APCs, e.g.,][]{Pinna_et_al_2010}.

In the following section, the model formulation is given in a generalisable form, yet
  the nomenclature is shifted towards urban train network, for the example
  implementation presented in subsequent section focuses on the Boston subway system.

\section{Model formulation}

The model represents macroscopic dynamics of passenger loading
  in a transit network conceptualised as an arbitrary connected graph
  with $n_s$ station nodes and $n_e$ edges.
For every edge in the graph, there is a pair of state-variable triplets defined
  - one triplet for northbound and another for southbound directions,
  each consisting of ``boarding'', ``on-board'' and ``alighting'' passenger counts
  \citep[nomenclature consistent e.g., with][]{Nuzzolo_et_al_2016}.
The system state vector $x$ consists, thus, of $n_x=6 \times n_e$ state variables hereinafter referred to as grid boxes.

The system dynamics covering passenger flow (due to vehicle movement), response to control input
  (inflow and outflow of passenger to/from stations) and observation process
  (crowdedness perception feedback) are linear and represented in line with the
  stochastic formulation underlying the Kalman filter with control input:
\begin{eqnarray*}
  x_{_k} &\!\!\!\!=&\!\!\!\! F_{_k} x_{_{k-1}} + B_{_k} u_{_k} + w_{_k}\\
  z_{_k} &\!\!\!\!=&\!\!\!\! H x_{_k} + v_{_k}
\end{eqnarray*}
where $k$ denotes timestep index, $z$ is the vector of passenger count measurements, and the right-hand side terms are defined with:
\begin{itemize}
  \item{$F_{_k}$: state transition matrix for vehicle movements and transfer probabilities,}
  \item{$B_{_k}$: control transition matrix representing north-/southbound probabilities,}
  \item{$u_{_k}$: control input vector with entry/exit counts (turnstiles, smart cards, APCs),}
  \item{$H$: observation matrix expressing spatial granularity of measurements,}
  \item{$w_{_k}, v_{_k}$: Gaussian noise terms expressing model and observation uncertainties.}
\end{itemize}

In the case of a single line, $F_{_k}$ can be thought of as a representation of a Marey diagram using a Boolean bidiagonal
  time-dependent matrix with values of one set either on the diagonal (e.g., passengers waiting on
  boarding platforms) or on the subdiagonal (passengers moving to adjacent grid boxes).
Flow of passengers between adjacent grid boxes is associated with vehicle departure or arrival
  at a station, which must occur in different time steps to account for
  movement of passengers between ``alighting'' and ``boarding'' grid boxes of a station
  (allowing also for representation of actual dwell times).
The rationale of introducing ``alighting'' and ``boarding'' grid boxes is to allow
  for representation of probabilities of transfers in multi-line networks
  (in which case the $F_{_k}$ matrix is no longer bidiagonal or Boolean).

The control vector $u_{_k}$ is meant to contain estimates of passenger inflow and outflow at
  station entry and exit gates expressed in passenger count per timestep.
The role of $B_{_k}$ is to represent the probabilities of passengers travelling
  north- or southbound (both in the case of arrival or departure).
The matrix could, thus, be derived from an origin-destination (OD) matrix or more simply
  by defining a gravity-like model in which $B$ is constant in time and
  conveys the assumption that in the centre of gravity (centre of the city)
  the probability of north- or south-bound travel is 50\%/50\%, at termini
  it is 100\%/0\% and remaining matrix coefficients are linearly interpolated.
In a broader context, the $Bu$ term allows to feed the model with differential macroscopic data, i.e.
  the numbers of passenger added or subtracted at a given time, at a given location
  in the network.

The interplay between the two additive terms $F_{_k}x_{_{k-1}}$ and $B_{_k}u_{_k}$
  makes the model capable of representing, e.g., accumulation of waiting passengers on
  platforms.
At the same time, it has to be noted that the interplay between these two terms may lead
  to appearance of negative passenger loading.
In the case of an alighting grid box, it can be interpreted as an expression of
  station's ``potential'' as a destination.
In the case of a vehicle carrying fewer passengers than expected to alight at
  a station, the negative signal will propagate to adjacent stations.
Negative values in ``on-board'' grid boxes are proposed to
  trigger feedback requests aimed at correcting the bogus datum after assimilating
  the feedback response.

The role of the observation process is to enable assimilation of integral
  macroscopic data, i.e. the numbers of passengers present at a given time at a given
  location in the network.
The $H$ matrix allows to express the spatial extent of the measurement, e.g. that a platform-reported crowdedness is to
  be interpreted as the sum of both northbound and southbound passenger counts.
Similarly, a summation over consecutive ``on-board'' grid boxes encoded in $H$ can express
  positioning uncertainty in correlating feedback from on-the-move passengers with a particular grid box.
Noteworthy, $H$ and $z$ can embody data and assumptions pertinent to multiple measurement
  methods.

Presence of the noise terms that embody model and feedback data uncertainties allows to
  apply the standard Kalman optimal filtering scheme widely used in transportation applications \citep[see e.g.,][section III and references therein]{Pagani_et_al_2016}. 
It is noteworthy, that given the Markovian character of the model and the central role of nearest-neighbour
  interactions (adjacent grid boxes), the model bares some resemblance
  with a scalar conservation problem -- with $B_{_k}u_{_k}$ resembling
  a source term and the ones on the subdiagonal of $F_{_k}$ corresponding to Courant numbers of
  unity, using the PDE numerics nomenclature.
The predict step of the Kalman filter is bound to a conservation constraint
  $\Sigma u_k + \Sigma x_{k-1} = \Sigma x_k$ (summations over all elements of $u$ and $x$)
  expressing conservation of total passenger count in the network and constituting a
  validity condition for $F_k$ and $B_k$.

\section{Prototype implementation and sample results}

\begin{figure}[t]
  \center
  \includegraphics[width=\textwidth]{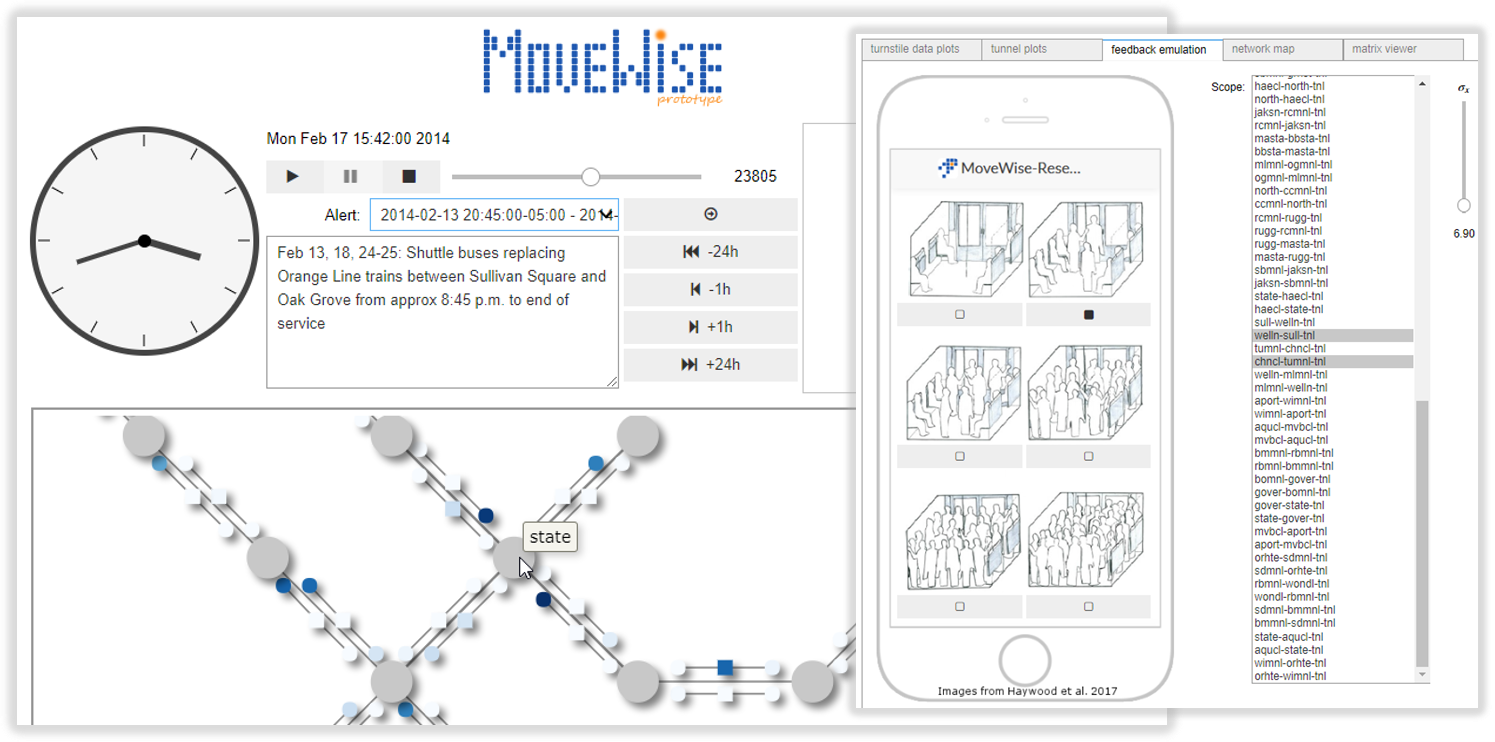}
  \caption{\label{fig:interface}
     Screenshot depicting selected components of the MoveWise prototype interface
  }
\end{figure}

The above-defined representation of the network and its dynamics has been
  prototypically implemented in Python using Jupyter, networkX, 
  FilterPy \citep{Labbe_2018} and visJS2jupyter \citep{Rosenthal_et_a_2017}.
The prototype has been applied to simulate passenger loading in the Boston
  subway network using an open dataset released with the MBTAViz project 
  of Barry and Card ({http://mbtaviz.github.io/}).
The dataset consists of four-week-long timeseries of turnstile inflow and outflow 
  passenger counts with 1-minute time resolution, of the train departure and
  arrival times and of a record of alerts on service disruptions.
The dataset covers three lines: red, orange and blue 
  \citep[as compared to red-line-limited datasets used in][]{Heimburger_et_al_1999, Koutsopoulos_and_Wang_2007}.
As a side note, interestingly, the Boston subway has been used as an example
  in studies of the so-called small-world networks
  \citep{Marchiori_and_Latora_2000, Latora_and_Marchiori_2002}.

Since the original dataset has a mixed microscopic/macroscopic character with the 
  train arrival/departure times associated with particular vehicles,
  these data were aggregated using the resolution of the turnstile dataset 
  and recast in a macroscopic form of counts of trains departing and arriving at a given station
  \citep[effectively carrying the same information content as the vehicle events defined in][]{Reyes_and_Cipriano_2014}.

Figure~\ref{fig:interface} depicts selected components of the interface
  of the developed prototype implementation named MoveWise.
The top left section of the figure depicts simulation controls not discussed herein.
The bottom left section of the figure depicts the interactive animated 
  representation of the subway network and simulation grid with
  passenger loading indicated with a colour scale.
The ``boarding'' and ``alighting'' grid boxes are represented with small circles,
  ``on-board'' ones with squares.
The right section of the figure depicts the feedback emulation panel with
  a smartphone mock-up and an application interface consisting of
  six cartoons corresponding to different crowdedness levels 
  \citep[images adapted from][]{Haywood_et_al_2017}.

\begin{figure}
  \center
  \includegraphics[width=.495\textwidth]{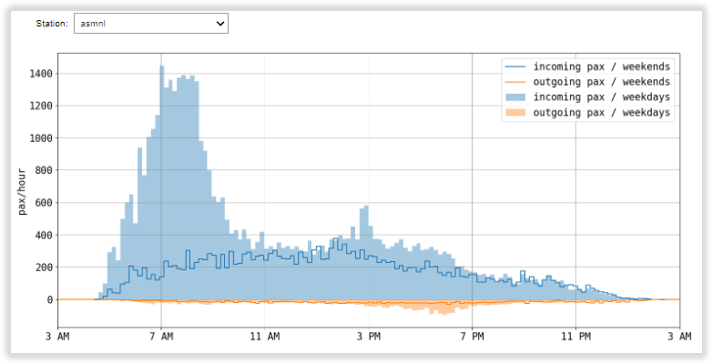}
  \includegraphics[width=.495\textwidth]{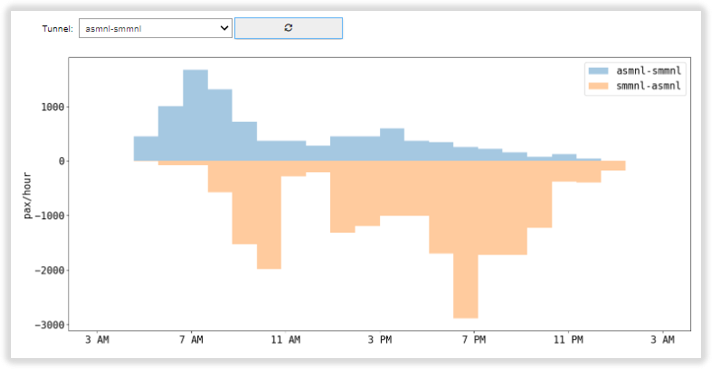}
  \caption{\label{fig:sample}
    {\bf Left panel:} mean values of passenger counts at the Ashmont terminus based
      on the four-week-long timeseries of turnstile data, computed separately for 
      weekdays (filled histogram) and weekends (line plot), binned into 10-minute
      intervals and plotted as a function of time of day.
    {\bf Right panel:} amounts of passengers moving from ``boarding''
      to ``on-board'' grid points between the Ashmont terminus and the
      next station, Shawmut, in a simulation driven by data from Feb 25 2014, 
      binned into 60-min intervals and plotted with colours corresponding to
      the left panel plot -- see text for discussion.
  }
\end{figure}

Two panels of Figure~\ref{fig:sample} depict input and output data of a
  sample simulation.
The plot in the left panel reveals documented deficiency of the dataset, namely
  missing (or largely underestimated) data on the numbers of passengers exiting 
  from some of the stations.
Since the depicted data correspond to a terminus (Ashmont station), 
  the missing data
  on the amounts of passengers exiting the station ought to be retrievable from
  simulation output, for all passengers arriving at the terminus are
  assumed to exit the station.
The plot in the right panel confirms plausibility of this hypothesis:
  the depicted breakdown of passengers travelling from the penultimate
  station to the terminus has an arguably realistic time profile.
The magnitude of the outgoing passenger flux (orange filled histogram)
  is likely overestimated, for any errors in the turnstile data
  (in particular, uncounted exiting passengers) accumulate along a line 
  and cause overestimation of simulated passenger counts at the terminus.
Presented simulation was performed without any feedback data and with the simple gravity-like
  assumption resulting in constant-in-time matrix $B$ and with an 
  arbitrary transfer probabilities at the fork of the red line (50\%-50\% 
  probabilities of transferring towards Ashmont or Braintree for passengers going out of town, 95\% probability to
  go towards the centre of town for passengers arriving at the JFK fork from the south).
Presented output data exemplifies that, despite these simplifications, the
  methodology is capable of enriching (or gap-filling) existing datasets,
  even without leveraging the potential for assimilation of crowdedness
  measurements.
  
\section{Summary and discussion}

\vspace{-.4em}
This note covered discussion of an approach to modelling passenger number dynamics
  (and conservation) within a transit network using the system of equations 
  congruent with the underlying equations of the Kalman filter with control input.
Presented approach allows to assimilate both integral (passenger counts)
  and differential (passenger fluxes)
  data on passenger density, along with vehicle movement data.
Sample results from a simulation driven by  
  vehicle-location and passenger-flow data for the Boston subway 
  exemplify how the methodology can be used to enrich an existing dataset.

The potential of the methodology lies in leveraging the applicable
  filtering techniques (Kalman filter and beyond) for
  setting up data fusion workflows feeding from diverse sources
  of data on passenger density and assimilating them taking into
  account the relevant information on spatial and temporal
  uncertainty of the data.
One of the advantages of the methodology is the potential for
  rapid implementation thanks to employment of a standard model
  for which software libraries are readily available.

This note outlined an idea of a participatory crowdsensing platform focused on real-time provision
  of information on the levels of crowding.
In such scenario, the role of the presented state-space model and the data 
  assimilation methodology is twofold: (i) to assimilate passenger-reported
  feedback into simulations driven by operator-provided data and (ii)
  to provide basis for defining feedback-request triggers using the 
  estimated system state as well as using the filter performance indices.

\vspace{-1.4em}
\section*{Acknowledgements}

The authors wish to thank Eleni I. Vlahogianni (National Technical University of~Athens) and with Alexandros Deloukas
(ATTIKO METRO) for helpful discussions. 

\vspace{-1.4em}
\bibliographystyle{tfcad}
\bibliography{paper}

\end{document}